\documentstyle[aps,preprint,eqsecnum,epsf]{revtex}

\tightenlines

\newcommand{\bq}{\begin{quotation}\noindent}
\newcommand{\eq}{\end{quotation}\medskip}

\def\mat#1#2#3#4{\left(\matrix{{#1}&{#2}\cr {#3}&{#4}}\right)}
\def\tr{{\rm tr}}

\def\vec#1{{\bf #1}}

\begin{document}

\draft

\title{Unknown Quantum States: \\
The Quantum de Finetti Representation}

\author{Carlton M. Caves,$^1$ Christopher A. Fuchs,$^2$ and
R\"udiger Schack$^3$\medskip}

\address{$^1$Department of Physics and Astronomy, University of New Mexico,
\\
Albuquerque, New Mexico 87131--1156, USA
\\
$^2$Computing Science Research Center, Bell Labs, Lucent Technologies,
\\
Room 2C-420, 600--700 Mountain Avenue, Murray Hill, New Jersey 07974,
USA
\\
$^3$Department of Mathematics, Royal Holloway, University of London,
\\
Egham, Surrey TW20$\;$0EX, UK}

\date{17 March 2001}

\maketitle

\begin{abstract}
We present an elementary proof of the {\it quantum de Finetti
representation theorem}, a quantum analogue of de Finetti's
classical theorem on exchangeable probability assignments.  This
contrasts with the original proof of Hudson and Moody [Z.\
Wahrschein.\ verw.\ Geb.\ {\bf 33}, 343 (1976)], which relies on
advanced mathematics and does not share the same potential for
generalization.  The classical de Finetti theorem provides an
operational definition of the concept of an unknown probability in
Bayesian probability theory, where probabilities are taken to be
degrees of belief instead of objective states of nature.  The
quantum de Finetti theorem, in a closely analogous fashion, deals
with exchangeable density-operator assignments and provides an
operational definition of the concept of an ``unknown quantum
state'' in quantum-state tomography.  This result is especially
important for information-based interpretations of quantum
mechanics, where quantum states, like probabilities, are taken to
be states of knowledge rather than states of nature.  We further
demonstrate that the theorem fails for real Hilbert spaces and
discuss the significance of this point.
\end{abstract}

\section{Introduction} \label{sec-intro}

What is a quantum state?  Since the earliest days of quantum theory,
the predominant answer has been that the quantum state is a
representation of the observer's knowledge of a
system~\cite{Bohr1928}.  In and of itself, the quantum state has no
objective reality~\cite{Fuchs2000}.  The authors hold this
information-based view quite firmly~\cite{Caves1996,Caves1997}.
Despite its association with the founders of quantum theory,
however, holding this view does not require a concomitant belief
that there is nothing left to learn in quantum foundations. It is
quite the opposite in fact: Only by pursuing a promising, but
incomplete program can one hope to learn something of lasting value.
Challenges to the information-based view arise regularly, and
dealing with these challenges builds an understanding and a
problem-solving agility that reading and rereading the founders can
never engender~\cite{Faye1994}. With each challenge successfully
resolved, one walks away with a deeper sense of the physical content
of quantum theory and a growing confidence for tackling questions of
its interpretation and applicability. Questions as fundamental and
distinct as ``Will a nonlinear extension of quantum mechanics be
needed to quantize gravity?''~\cite{tHooft1999,Jozsa1998} and
``Which physical resources actually make quantum computation
efficient?''~\cite{Schack1999,Ambainis2000} start to feel tractable
(and even connected) from this perspective.

In this paper, we tackle an understanding-building exercise very
much in the spirit of these remarks.  It is motivated by an
apparent conundrum arising from our own specialization in physics,
quantum information theory.  The issue is that of the {\it
unknown\/} quantum state.

There is hardly a paper in the field of quantum information that
does not make use of the idea of an ``unknown quantum state.''
Unknown quantum states are
teleported~\cite{Bennett1993,Experiments1998}, protected with
quantum error correcting codes~\cite{Shor1995,Steane1996}, and used
to check for quantum
eavesdropping~\cite{Bennett1984,CryptoExperiments}.  The list of
uses, already long, grows longer each day.  Yet what can the term
``unknown quantum state'' mean? In an information-based
interpretation of quantum mechanics, the term is an oxymoron:  If
quantum states, by their very definition, are states of knowledge
and not states of nature~\cite{Hartle1968}, then the state is {\it
known\/} by someone---at the very least, by the describer himself.

This message is the main point of our paper.
Faced with a procedure that uses the
idea of an unknown quantum state in its description, a consistent
information-based interpretation of quantum mechanics offers only
two alternatives:
\begin{itemize}
\item
The owner of the unknown state---a further decision-making agent or
observer---must be explicitly identified.  In this case, the unknown
state is merely a stand-in for the unknown {\it state of
knowledge\/} of an essential player who went unrecognized in the
original formulation.
\item
If there is clearly no further decision-making agent or observer
on the scene, then a way must be found to re\"express the
procedure with the term ``unknown state'' banished from the
formulation. In this case, the end-product of the effort is a
single quantum state used for describing the entire
procedure---namely, the state that captures the describer's state
of knowledge.
\end{itemize}

Of course, those inclined to an objectivist interpretation of
quantum mechanics~\cite{Goldstein98}---that is, an interpretation
where quantum states are more like states of nature than states of
knowledge---might be tempted to believe that the scarcity of
existing analyses of this kind is a hint that quantum states do
indeed have some sort of objective status.  Why would such currency
be made of the unknown-state concept were it not absolutely
necessary?  As a rejoinder, we advise caution to the objectivist:
Tempting though it is to grant objective status to all the
mathematical objects in a physical theory, there is much to be
gained by a careful delineation of the subjective and objective
parts. A case in point is provided by E.~T. Jaynes'
\cite{Jaynes1957a,Jaynes1957b,Jaynes1983} insistence that entropy is
a subjective quantity, a measure of ignorance about a physical
system.  One of the many fruits of this point of view can be found in
the definitive solution \cite{Bennett1983} to the long-standing
Maxwell demon problem \cite{Leff1990}, where it was realized that the
information collected by a demon and used by it to extract work from
heat has a thermodynamic cost at least as large as the work extracted
\cite{Landauer1961}.

\begin{center}
\begin{figure} \leavevmode
\epsfxsize=8cm \epsfbox{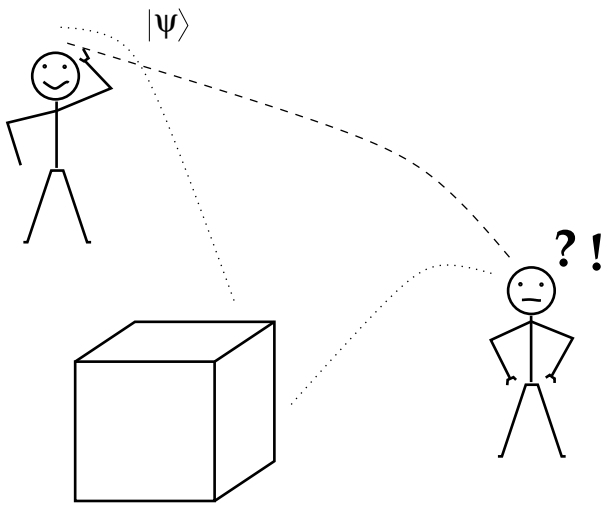} \bigskip\caption{What can the term
``unknown state'' mean if quantum states are taken exclusively to be
states of knowledge rather than states of nature?  When we say that
a system has an unknown state, must we always imagine a further
observer whose state of knowledge is symbolized by some
$|\psi\rangle$, and it is the identity of the symbol that we are
ignorant of?}
\end{figure}
\end{center}

The example analyzed in detail in this paper provides another case.
Along the way, it brings to light a new and distinct point about why
quantum mechanics makes use of complex Hilbert spaces rather than
real or quaternionic
ones~\cite{Stueckelberg1960,Adler1995,Araki1980,Wootters1990}.
Furthermore, the method we use to prove our main theorem employs a
novel measurement technique that might be of use in the laboratory.

We analyze in depth a particular use of unknown states, which
comes from the measurement technique known as {\it quantum-state
tomography\/}~\cite{Vogel1989b,Smithey1993,Leonhardt1995}. The
usual description of tomography is this.  A device of some sort,
say a nonlinear optical medium driven by a laser, repeatedly
prepares many instances of a quantum system, say many temporally
distinct modes of the electromagnetic field, in a fixed quantum
state $\rho$, pure or mixed.  An experimentalist who wishes to
characterize the operation of the device or to calibrate it for
future use might be able to perform measurements on the systems it
prepares even if he cannot get at the device itself.  This can be
useful if the experimenter has some prior knowledge of the
device's operation that can be translated into a probability
distribution over states. Then learning about the state will also
be learning about the device.  Most importantly, though, this
description of tomography assumes that the precise state $\rho$ is
unknown.  The goal of the experimenter is to perform enough
measurements, and enough kinds of measurements (on a large enough
sample), to estimate the identity of $\rho$.

This is clearly an example where there is no further player on
whom to pin the unknown state as a state of knowledge.  Any
attempt to find a player for the pin is entirely artificial: Where
would the player be placed?  On the inside of the device the
tomographer is trying to characterize \cite{BerkeleyRhyme}?  The
only available course for an information-based interpretation of
quantum-state tomography is the second strategy listed above---to
banish completely the idea of the unknown state from the formulation
of tomography.

\begin{center}
\begin{figure} \leavevmode
\epsfxsize=9cm \epsfbox{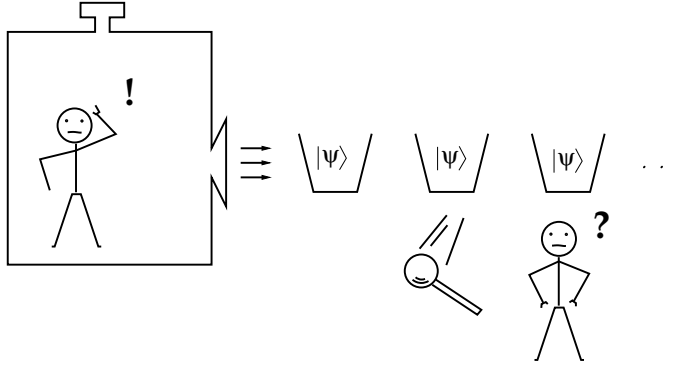} \bigskip\caption{To make sense of
quantum tomography, must we go to the extreme of imagining a ``man
in the box'' who has a better description of the systems than we
do?  How contrived our usage would be if that were so!}
\end{figure}
\end{center}

To do this, we take a cue from the field of Bayesian probability
theory~\cite{Kyburg1980,JaynesPosthumous,Bernardo1994}, prompted
by the realization that Bayesian probability is to probability
theory in general what an information-based interpretation is to
quantum mechanics~\cite{Caves1996,Schack1997}. In Bayesian theory,
probabilities are not objective states of nature, but rather are
taken explicitly to be measures of credible belief, reflecting
one's state of knowledge.  The overarching Bayesian theme is to
identify the conditions under which a set of decision-making
agents can come to a common belief or probability assignment for a
random variable even though their initial beliefs
differ~\cite{Bernardo1994}.  Following that theme is the key to
understanding tomography from the informational point of view.

The offending classical concept is an ``unknown probability,'' an
oxymoron for the same reason as an unknown quantum state.  The
procedure analogous to quantum-state tomography is the estimation of
an unknown probability from the results of repeated trials on
``identically prepared systems,'' all of which are said to be
described by the same, but unknown probability.  The way to
eliminate unknown probabilities from the discussion, introduced by
Bruno de Finetti in the early 1930s
\cite{DeFinetti1990,DeFinettiCollected}, is to focus on the
equivalence of repeated trials, which means the systems are
indistinguishable as far as probabilistic predictions are concerned
and thus that a probability assignment for multiple trials should be
symmetric under permutation of the systems.  With his {\it classical
representation theorem}, de Finetti \cite{DeFinetti1990} showed that
a multi-trial probability assignment that is permutation-symmetric
for an arbitrarily large number of trials---de Finetti called such
multi-trial probabilities {\it exchangeable\/}---is equivalent to a
probability for the ``unknown probabilities.''  Thus the
unsatisfactory concept of an unknown probability vanishes from the
description in favor of the fundamental idea of assigning an
exchangeable probability distribution to multiple trials.

This cue in hand, it is easy to see how to reword the description
of quantum-state tomography to meet our goals.  What is relevant
is simply a judgment on the part of the experimenter---notice the
essential subjective character of this ``judgment''---that there
is no distinction between the systems the device is preparing.  In
operational terms, this is the judgment that {\it all the systems
are and will be the same as far as observational predictions are
concerned}.  At first glance this statement might seem to be
contentless, but the important point is this: To make this
statement, one need never use the notion of an unknown state---a
completely operational description is good enough. Putting it into
technical terms, the statement is that if the experimenter judges
a collection of $N$ of the device's outputs to have an overall
quantum state $\rho^{(N)}$, he will also judge any permutation of
those outputs to have the same quantum state $\rho^{(N)}$.
Moreover, he will do this no matter how large the number $N$ is.
This, complemented only by the consistency condition that for any
$N$ the state $\rho^{(N)}$ be derivable from $\rho^{(N+1)}$, makes
for the complete story.

The words ``quantum state'' appear in this formulation, just as in
the original formulation of tomography, but there is no longer any
mention  of {\it unknown\/} quantum states.  The state $\rho^{(N)}$
is known by the experimenter (if no one else), for it represents his
state of knowledge.  More importantly, the experimenter is in a
position to make an unambiguous statement about the structure of the
whole sequence of states $\rho^{(N)}$: Each of the states
$\rho^{(N)}$ has a kind of permutation invariance over its factors.
The content of the {\it quantum de Finetti representation
theorem}~\cite{Hudson1976,Hudson1981}---a new proof of which is the
main technical result of this paper---is that a sequence of states
$\rho^{(N)}$ can have these properties, which are said to make it an
{\it exchangeable\/} sequence, if and only if each term in it can
also be written in the form
\begin{equation}
\rho^{(N)}=\int P(\rho)\, \rho^{\otimes N}\, d\rho\;,
\label{Jeremy}
\end{equation}
where
\begin{equation}
\rho^{\otimes N}=
\underbrace{\rho\otimes\rho\otimes\cdots\otimes\rho}_{
\matrix{\mbox{$N$-fold tensor}\cr\mbox{product}}}
\end{equation}
and $P(\rho)$ is a fixed probability distribution over the density
operators.

The interpretive import of this theorem is paramount. It alone
gives a mandate to the term unknown state in the usual description
of tomography.  It says that the experimenter can act {\it as
if\/} his state of knowledge $\rho^{(N)}$ comes about because he
knows there is a ``man in the box,'' hidden from view, repeatedly
preparing the same state $\rho$.  He does not know which such
state, and the best he can say about the unknown state is captured
in the probability distribution $P(\rho)$.

The quantum de Finetti theorem furthermore makes a connection to the
overarching theme of Bayesianism stressed above.  It guarantees for
two independent observers---as long as they have a rather minimal
agreement in their initial beliefs---that the outcomes of a
sufficiently informative set of measurements will force a
convergence in their state assignments for the remaining
systems~\cite{Schack2000}.  This ``minimal'' agreement is
characterized by a judgment on the part of both parties that the
sequence of systems is exchangeable, as described above, and a
promise that the observers are not absolutely inflexible in their
opinions.  Quantitatively, the latter means that though $P(\rho)$
might be arbitrarily close to zero, it can never vanish.

This coming to agreement works because an exchangeable density
operator sequence can be updated to reflect information gathered
from measurements by a quantum version of Bayes's rule for
updating probabilities. Specifically, if measurements on $K$
systems yield results $D_K$, then the state of additional systems
is constructed as in Eq.~(\ref{Jeremy}), but using an updated
probability on density operators given by
\begin{equation}
P(\rho|D_K)={P(D_K|\rho)P(\rho)\over P(D_K)}\;.
\label{QBayes}
\end{equation}
Here $P(D_K|\rho)$ is the probability to obtain the measurement
results $D_K$, given the state $\rho^{\otimes K}$ for the $K$
measured systems, and $P(D_K)=\int P(D_K|\rho)\,P(\rho)\,d\rho$ is
the unconditional probability for the measurement results.
Equation~(\ref{QBayes}) is a kind of {\it quantum Bayes rule}
\cite{Schack2000}.  For a sufficiently informative set of
measurements, as $K$ becomes large, the updated probability
$P(\rho|D_K)$ becomes highly peaked on a particular state
$\rho_{D_K}$ dictated by the measurement results, regardless of the
prior probability $P(\rho)$, as long as $P(\rho)$ is nonzero in a
neighborhood of $\rho_{D_K}$.  Suppose the two observers have
different initial beliefs, encapsulated in different priors
$P_i(\rho)$, $i=1,2$.  The measurement results force them to a
common state of knowledge in which any number $N$ of additional
systems are assigned the product state $\rho_{D_K}^{\otimes N}$,
i.e.,
\begin{equation}
\int P_i(\rho|D_K)\,\rho^{\otimes N}\,d\rho
\quad{\longrightarrow}\quad
\rho_{D_K}^{\otimes N}\;,
\label{HannibalLecter}
\end{equation}
independent of $i$, for $K$ sufficiently large.

This shifts the perspective on the purpose of quantum-state
tomography:  It is not about uncovering some ``unknown state of
nature,'' but rather about the various observers' coming to
agreement over future probabilistic predictions~\cite{Fuchs2000b}.
In this connection, it is interesting to note that the quantum de
Finetti theorem and the conclusions just drawn from it work only
within the framework of complex vector-space quantum mechanics.
For quantum mechanics based on real and quaternionic Hilbert
spaces~\cite{Stueckelberg1960,Adler1995}, the connection between
exchangeable density operators and unknown quantum states does
not hold.

The plan of the remainder of the paper is as follows. In
Sec.~\ref{sec-classical} we discuss the classical de Finetti
representation theorem~\cite{DeFinetti1990,Heath1976} in the
context of Bayesian probability theory.  It was our familiarity
with the classical theorem~\cite{Galavotti1989,Jeffrey1997} that
motivated our reconsideration of quantum-state tomography. In
Sec.~\ref{sec-quantum} we introduce the information-based
formulation of tomography in terms of exchangeable multi-system
density operators, accompanied by a critical discussion of
objectivist formulations of tomography, and we state the quantum
de Finetti representation theorem.  Section~\ref{sec-proof}
presents an elementary proof of the quantum de Finetti theorem.
There, also, we introduce a novel measurement technique for
tomography based upon generalized quantum measurements.  Finally,
in Sec.~\ref{sec-outlook} we return to the issue of number fields
in quantum mechanics and mention possible extensions of the main
theorem.

\section{The Classical de Finetti Theorem} \label{sec-classical}

As a preliminary to the quantum problem, we turn our attention to
classical probability theory.   In doing so we follow a maxim of
the late E.~T. Jaynes~\cite{Jaynes1986b}:
\begin{quote}
We think it unlikely that the role of probability in quantum theory
will be understood until it is generally understood in classical theory
\ldots.  Indeed, our [seventy-five-year-old] bemusement over the
notion of state reduction in [quantum theory] need not surprise us
when we note that today, in all applications of probability theory,
basically the same controversy rages over whether our probabilities
represent real situations, or only incomplete human knowledge.
\end{quote}

As Jaynes makes clear, the tension between the objectivist and
informational points of view is not new with quantum mechanics.
It arises already in classical probability theory in the form of
the war between ``objective'' and ``subjective''
interpretations~\cite{Daston1994}. According to the subjective or
Bayesian interpretation, probabilities are measures of credible
belief, reflecting an agent's potential states of knowledge. On
the other hand, the objective interpretations---in all their
varied forms, from frequency interpretations to propensity
interpretations---attempt to view probabilities as real states of
affairs or ``states of nature.'' Following our discussion in
Sec.~\ref{sec-intro}, it will come as no surprise to the reader
that the authors wholeheartedly adopt the Bayesian approach. For
us, the ultimate reason is simply our own experience with this
question, part of which is an appreciation that objective
interpretations inevitably run into insurmountable difficulties.
We will not dwell upon these difficulties here; instead, the
reader can find a sampling of criticisms in
Refs.~\cite{Jaynes1983,Kyburg1980,JaynesPosthumous,Bernardo1994,%
Savage1972}.

We will note briefly, however, that the game of roulette provides
an illuminating example. In the European version of the game, the
possible outcomes are the numbers $0,1,\ldots,36$.  For a player
without any privileged information, all 37 outcomes have the same
probability $p=1/37$.  But suppose that shortly after the ball is
launched by the croupier, another player obtains information about
the ball's position and velocity relative to the wheel. Using the
information obtained, this other player can make more accurate
predictions than the first~\cite{NewtonianCasino}. His probability
is peaked around some group of numbers. The probabilities are
thus different for two players with different states of knowledge.

Whose probability is the true probability? From the Bayesian
viewpoint, this question is meaningless:  There is no such thing
as a true probability.  All probability assignments are subjective
assignments based specifically upon one's prior information.

For sufficiently precise data---including precise initial data on
positions and velocities and probably also including other details
such as surface properties of the wheel---Newtonian mechanics
assures us that the outcome can be predicted with certainty.  This
is an important point: The determinism of classical physics
provides a strong reason for adopting the subjectivist view of
probabilities~\cite{Giere1973}.  If the conditions of a trial are
exactly specified, the outcomes are predictable with certainty,
and all probabilities are 0 or 1. In a deterministic theory, all
probabilities strictly greater than 0 and less than 1 arise as a
consequence of incomplete information and depend upon their
assigner's state of knowledge.

Of course, we should keep in mind that our ultimate goal is to
consider the status of quantum states and, by way of them, quantum
probabilities. One can ask, ``Does this not change the flavor of
these considerations?'' Since quantum mechanics is avowedly {\it
not\/} a theory of one's ignorance of a set of hidden
variables~\cite{BellBook,GoldsteinBook}, how can the probabilities
be subjective?  In Sec.~\ref{sec-quantum} we argue that despite
the intrinsic indeterminism of quantum mechanics, the essence of
the point above carries over to the quantum setting intact.
Furthermore, there are specifically quantum-motivated arguments
for a Bayesian interpretation of quantum probabilities.

For the present, though, let us consider in some detail the
general problem of a repeated experiment---spinning a roulette
wheel $N$ times is an example. As discussed briefly in
Sec.~\ref{sec-intro}, this allows us to make a conceptual
connection to quantum-state tomography.  Here the individual
trials are described by discrete random variables
$x_n\in\{1,2,\ldots,k\}$, $n=1,\ldots,N$; that is to say, there
are $N$ random variables, each of which can assume $k$ discrete
values. In an objectivist theory, such an experiment has a
standard formulation in which the probability in the multi-trial
hypothesis space is given by an independent, identically
distributed (i.i.d.)\ distribution
\begin{equation}
p(x_1,x_2,\ldots,x_N)\,=\,p_{x_1} p_{x_2} \cdots p_{x_N}\, =\,
p_1^{n_{\scriptscriptstyle 1}} p_2^{n_{\scriptscriptstyle 2}}\cdots
p_k^{n_{\scriptscriptstyle k}}\;.
\label{eq-iid}
\end{equation}
The number $p_j$ ($j=1,\ldots,k$) describes the objective,
``true'' probability that the result of a single experiment will
be $j$ ($j=1,\ldots,k$).  The variable $n_j$, on the other hand,
is the number of times outcome $j$ is listed in the vector
$(x_1,x_2,\ldots,x_N)$.  This simple description---for the
objectivist---only describes the situation from a kind of ``God's
eye'' point of view.  To the experimentalist, the ``true''
probabilities $p_1,\ldots,p_k$ will very often be {\it unknown\/}
at the outset.  Thus, his burden is to estimate the unknown
probabilities by a statistical analysis of the experiment's
outcomes.

In the Bayesian approach, it does not make sense to talk about
estimating a true probability.  Instead, a Bayesian assigns a
prior probability distribution $p(x_1,x_2,\ldots,x_N)$ on the
multi-trial hypothesis space, which is generally not an i.i.d.,
and then uses Bayes's theorem to update the distribution in the
light of measurement results. A common criticism from the
objectivist camp is that the choice of distribution
$p(x_1,x_2,\ldots,x_N)$ with which to start the process seems
overly arbitrary to them. On what can it be grounded, they would
ask? From the Bayesian viewpoint, the subjectivity of the prior is
a strength rather than a weakness, because assigning a prior
amounts to laying bare the necessarily subjective assumptions
behind {\it any\/} probabilistic argument, be it Bayesian or
objectivist. Choosing a prior among all possible distributions on
the multi-trial hypothesis space is, however, a daunting task. As
we will now see, the de Finetti representation theorem makes this
task tractable.

It is very often the case that one or more features of a problem
stand out so clearly that there is no question about how to
incorporate them into an initial assignment. In the present case,
the key feature is contained in the assumption that an arbitrary
number of repeated trials are equivalent.  This means that one has
no reason to believe there will be a difference between one trial
and the next. In this case, the prior distribution is judged to
have the sort of permutation symmetry discussed briefly in
Sec.~\ref{sec-intro}, which de Finetti \cite{DeFinettiCollected}
called {\it exchangeability}.  The rigorous definition of
exchangeability proceeds in two stages.

A probability distribution $p(x_1,x_2,\ldots,x_N)$ is said to be
{\it symmetric\/} (or finitely exchangeable) if it is invariant
under permutations of its arguments, i.e., if
\begin{equation}
p\bigl(x_{\pi(1)},x_{\pi(2)},\ldots,x_{\pi(N)}\bigr) =
p(x_1,x_2,\ldots,x_N)
\end{equation}
for any permutation $\pi$ of the set $\{1,\ldots,N\}$. The
distribution $p(x_1,x_2,\ldots,x_N)$ is called {\it
exchangeable\/} (or infinitely exchangeable) if it is symmetric
and if for any integer $M>0$, there is a symmetric distribution
$p_{N+M}(x_1,x_2,\ldots,x_{N+M})$ such that
\begin{equation}
p(x_1,x_2,\ldots,x_N)\; =
\sum_{x_{N+1},\ldots,x_{N+M}}
p_{N+M}(x_1,\ldots,x_N,x_{N+1},\ldots,x_{N+M})
\;.
\label{eq-marginal}
\end{equation}
This last statement means the distribution $p$ can be extended to
a symmetric distribution of arbitrarily many random variables.
Expressed informally, an exchangeable distribution can be thought
of as arising from an infinite sequence of random variables whose
order is irrelevant.

We now come to the main statement of this section:
if a probability distribution $p(x_1,x_2,\ldots,x_N)$ is exchangeable,
then it can be written uniquely in the form
\begin{equation}
p(x_1,x_2,\ldots,x_N)=\int_{{\cal S}_k} P(\vec{p})\,p_{x_1}
p_{x_2} \cdots p_{x_N}\,d\vec{p}
=
\int_{{\cal S}_k} P(\vec{p})\, p_1^{n_{\scriptscriptstyle 1}}
p_2^{n_{\scriptscriptstyle 2}}\cdots p_k^{n_{\scriptscriptstyle k}} \,
d\vec{p}\;,
\label{eq-repr}
\end{equation}
where $\vec{p}=(p_1,p_2,\ldots,p_k)$, and the integral is taken over
the probability simplex
\begin{equation}
{\cal S}_k=\left\{\vec{p}\mbox{ : }\; p_j\ge0\mbox{ for all } j\mbox{
and } \sum_{j=1}^k p_j=1\right\}.
\end{equation}
Furthermore, the function $P(\vec{p})\ge0$ is required to be a
probability density function on the simplex:
\begin{equation}
\int_{{\cal S}_k} P(\vec{p})\,d\vec{p}=1\;.
\end{equation}
Equation~(\ref{eq-repr}) comprises the classical de Finetti
representation theorem for discrete random variables. For
completeness and because it deserves to be more widely familiar in
the physics community, we give a simple proof (due to Heath and
Sudderth \cite{Heath1976}) of the representation theorem for the
binary random-variable case in an Appendix.

Let us reiterate the importance of this result for the present
considerations.  It says that an agent, making solely the
judgment of exchangeability for a sequence of random variables
$x_j$, can proceed {\it as if\/} his state of knowledge had
instead come about through ignorance of an {\it unknown}, but
objectively existent set of probabilities $\vec{p}$.  His precise
ignorance of $\vec{p}$ is captured by the ``probability on
probabilities'' $P(\vec{p})$.  This is in direct analogy to what
we desire of a solution to the problem of the unknown quantum
state in quantum-state tomography.

As a final note before finally addressing the quantum problem in
Sec.~\ref{sec-quantum}, we point out that both conditions in the
definition of exchangeability are crucial for the proof of the de
Finetti theorem.  In particular, there are probability
distributions $p(x_1,x_2,\ldots,x_N)$ that are symmetric, but not
exchangeable.  A simple example is the distribution $p(x_1,x_2)$
of two binary random variables $x_1,x_2\in\{0,1\}$,
\begin{eqnarray}
&& p(0,0) = p(1,1) = 0\;,
\label{HocusPocus}
\\
&& p(0,1) = p(1,0) = \frac{1}{2} \;.
\label{Hiroshima}
\end{eqnarray}
One can easily check that $p(x_1,x_2)$ cannot be written as the
marginal of a symmetric distribution of three variables, as in
Eq.~(\ref{eq-marginal}). Therefore it can have no representation
along the lines of Eq.~(\ref{eq-repr}).  (For an extended
discussion of this, see Ref.~\cite{Jaynes1986}.)  Indeed,
Eqs.~(\ref{HocusPocus}) and (\ref{Hiroshima}) characterize a
perfect ``anticorrelation'' of the two variables, in contrast to
the positive correlation implied by distributions of de Finetti
form.  The content of this point is that both conditions in the
definition of exchangeability (symmetry under interchange and
infinite extendibility) are required to ensure, in colloquial
terms, ``that the future will appear much as the past''
\cite{vonPlato1989}, rather than, say, the opposite of the past.

\section{The quantum de Finetti representation} \label{sec-quantum}

Let us now return to the problem of quantum-state tomography
described in Sec.~\ref{sec-intro}. In the objectivist formulation of
the problem, a device repeatedly prepares copies of a system in the
same quantum state $\rho$. This is generally a mixed-state density
operator on a Hilbert space ${\cal H}_d$ of $d$ dimensions. We call
the totality of such density operators ${\cal D}_d$.  The joint
quantum state of the $N$ systems prepared by the device is then
given by
\begin{equation}
\rho^{\otimes N}=\rho\otimes\rho\otimes\cdots\otimes\rho \;,
\end{equation}
the $N$-fold tensor product of $\rho$ with itself. This, of
course, is a very restricted example of a density operator on the
tensor-product Hilbert space ${\cal H}_d^{\otimes N}\equiv {\cal
H}_d\otimes\cdots\otimes{\cal H}_d$. The experimenter, who
performs quantum-state tomography, tries to determine $\rho$ as
precisely as possible. Depending upon the version of the argument,
$\rho$ is interpreted as the ``true'' state of each of the systems
or as a description of the ``true'' preparation procedure.

We have already articulated our dissatisfaction with this way of
stating the problem, but we give here a further sense of why both
interpretations above are untenable.  Let us deal first with the
version where $\rho$ is regarded as the true, objective state of
each of the systems.  In this discussion it is useful to consider
separately the cases of mixed and pure states $\rho$.  The
arguments against regarding mixed states as objective properties
of a quantum system are essentially the same as those against
regarding probabilities as objective. In analogy to the roulette
example given in the previous section, we can say that, whenever
an observer assigns a mixed state to a physical system, one can
think of another observer who assigns a different state based on
privileged information.

The quantum argument becomes yet more compelling if the apparently
nonlocal nature of quantum states is taken into consideration.
Consider two parties, $A$ and $B$, who are far apart in space, say
several light years apart. Each party possesses a spin-$1\over 2$
particle. Initially the joint state of the two particles is the
maximally entangled pure state
${1\over\sqrt2}(|0\rangle|0\rangle+|1\rangle|1\rangle)$.
Consequently, $A$ assigns the totally mixed state
${1\over2}(|0\rangle\langle0|+|1\rangle\langle1|)$ to her own
particle. Now $B$ makes a measurement on his particle, finds the
result 0, and assigns to $A$'s particle the pure state
$|0\rangle$. Is this now the ``true,'' objective state of $A$'s
particle? At what precise time does the objective state of $A$'s
particle change from totally mixed to pure?  If the answer is
``simultaneously with $B$'s measurement,'' then what frame of
reference should be used to determine simultaneity?  These
questions and potential paradoxes are avoided if states are
interpreted as states of knowledge. In our example, $A$ and $B$
have different states of knowledge and therefore assign different
states. For a detailed analysis of this example, see
Ref.~\cite{Peres-9906a}; for an experimental investigation see
Ref.~\cite{Scarani2000}.

If one admits that mixed states cannot be objective properties,
because another observer, possessing privileged information, can
know which pure state underlies the mixed state, then it becomes
very tempting to regard the pure states as giving the ``true''
state of a system.  Probabilities that come from pure states would
then be regarded as objective, and the probabilities for pure
states within an ensemble decomposition of a mixed state would be
regarded as subjective, expressing our ignorance of which pure
state is the ``true'' state of the system.  An immediate and, in
our view, irremediable problem with this idea is that a mixed
state has infinitely many ensemble decompositions into pure states
\cite{Jaynes1957b,Schrodinger1936,Hughston1993}, so the
distinction between subjective and objective becomes hopelessly
blurred.

This problem can be made concrete by the example of a spin-${1\over2}$
particle. Any pure state of the particle can be written in terms
of the Pauli matrices,
\begin{equation}  \label{eq-pauli}
\sigma_1={\mat0110}\;,\qquad \sigma_2={\mat0{-i}i0}\;,\qquad
\sigma_3={\mat100{-1}}\;,
\end{equation}
as
\begin{equation}  \label{eq-poincare}
|\vec n\rangle\langle\vec n|={1\over2}(I+{\vec n}\cdot\bbox{\sigma})
={1\over2}(I+n_1\sigma_1+n_2\sigma_2+n_3\sigma_3)\;,
\end{equation}
where the unit vector ${\vec n}=n_1\vec e_1+n_2\vec e_2+n_3\vec
e_3$ labels the pure state, and $I$ denotes the unit operator.  An
arbitrary state $\rho$, mixed or pure, of the particle can be
expressed as
\begin{equation}
\rho={1\over2}(I+\vec S\cdot\bbox{\sigma}) \;,
\label{eq-rhoqubit}
\end{equation}
where $0\le|\vec S|\le1$.  This representation of the states of a
spin-$1\over2$ particle is called the {\it Bloch-sphere
representation.} If $|\vec S|<1$, there is an infinite number of
ways in which $\vec S$ can be written in the form $\vec S=\sum_j
p_j{\vec n}_j$, $|\vec n_j|=1$, with the numbers $p_j$ comprising
a probability distribution, and hence an infinite number of
ensemble decompositions of $\rho$:
\begin{equation}
\rho = \sum_jp_j{1\over2}(I+{\vec n}_j\cdot\bbox{\sigma})
=\sum_j p_j|\vec n_j\rangle\langle\vec n_j|\;.
\label{eq-decomp}
\end{equation}

Suppose for specificity that the particle's state is a mixed state
with $\vec S={1\over2}\,\vec e_3$.  Writing $\vec S={3\over4}\vec
e_3+{1\over4}(-\vec e_3)$ gives the eigendecomposition,
\begin{equation}
\rho=
{3\over4}|\vec e_3\rangle\langle\vec e_3|
+{1\over4}|\mathord{-}\vec e_3\rangle\langle\mathord{-}\vec e_3|\;,
\end{equation}
where we are to regard the probabilities $3/4$ and $1/4$ as
subjective expressions of ignorance about which eigenstate is the
``true'' state of the particle.  Writing $\vec S={1\over2}\vec
n_++{1\over2}\vec n_-$, where $\vec n_{\pm}={1\over2}\vec
e_3\pm{\sqrt3\over2}\vec e_x$, gives another ensemble
decomposition,
\begin{equation}
\rho=
{1\over2}|\vec n_+\rangle\langle\vec n_+|
+{1\over2}|\vec n_-\rangle\langle\vec n_-|\;,
\label{Eleanor}
\end{equation}
where we are now to regard the two probabilities of $1/2$ as
expressing ignorance of whether the ``true'' state is $|\vec
n_+\rangle$ or $|\vec n_-\rangle$.

The problem becomes acute when we ask for the probability that a
measurement of the $z$ component of spin yields spin up; this
probability is given by $\langle\vec e_3|\rho|\vec
e_3\rangle={1\over2}(1+{1\over2}\langle\vec e_3|\sigma_3|\vec
e_3\rangle)=3/4$.  The eigendecomposition gets this probability by
the route
\begin{equation}
\langle\vec e_3|\rho|\vec e_3\rangle=
{3\over4}
\underbrace{|\langle\vec e_3|\vec e_3\rangle|^2}_%
{\displaystyle{1}}
+{1\over4}
\underbrace{|\langle\vec e_3|\mathord{-}\vec e_3\rangle|^2}_%
{\displaystyle{0}}\;.
\end{equation}
Here the ``objective'' quantum probabilities, calculated from the
eigenstates, report that the particle definitely has spin up or
definitely has spin down; the overall probability of $3/4$ comes
from mixing these objective probabilities with the subjective
probabilities for the eigenstates.  The
decomposition~(\ref{Eleanor}) gets the same overall probability by
a different route,
\begin{equation}
\langle\vec e_3|\rho|\vec e_3\rangle=
{1\over2}
\underbrace{|\langle\vec e_3|\vec n_+\rangle|^2}_%
{\displaystyle{3/4}}
+{1\over2}
\underbrace{|\langle\vec e_3|\vec n_-\rangle|^2}_%
{\displaystyle{3/4}}
\;.
\end{equation}
Now the quantum probabilities tell us that the ``objective''
probability for the particle to have spin up is $3/4$.  This
simple example illustrates the folly of trying to have two kinds
of probabilities in quantum mechanics.  The lesson is that if a
density operator is even partially a reflection of one's state of
knowledge, the multiplicity of ensemble decomposition means that a
pure state must also be a state of knowledge.

Return now to the second version of the objectivist formulation of
tomography, in which the experimenter is said to be using
quantum-state tomography to determine an unknown preparation
procedure. Imagine that the tomographic reconstruction results in
the mixed state $\rho$, rather than a pure state, as in fact all
actual laboratory procedures do.   Now there is a serious problem,
because a mixed state does not correspond to a well-defined
procedure, but is itself a probabilistic mixture of well-defined
procedures, i.e., pure states.  The experimenter is thus trying to
determine an unknown procedure that has no unique decomposition
into well defined procedures.  Thus he cannot be said to be
determining an unknown procedure at all.  This problem does not
arise in an information-based interpretation, according to which
all quantum states, pure or mixed, are states of knowledge.  In
analogy to the classical case, the quantum de Finetti
representation provides an operational definition for the idea of
an unknown quantum state in this case.

Let us therefore turn to the information-based formulation of the
quantum-state tomography problem. Before the tomographic
measurements, the Bayesian experimenter assigns a prior quantum
state to the joint system composed of the $N$ systems, reflecting
his prior state of knowledge.  Just as in the classical case, this
is a daunting task unless the assumption of exchangeability is
justified.

The definition of the quantum version of exchangeability is
closely analogous to the classical definition.  Again, the
definition proceeds in two stages.  First, a joint state
$\rho^{(N)}$ of $N$ systems is said to be {\it symmetric\/} (or
finitely exchangeable) if it is invariant under any permutation of
the systems.  To see what this means formally, first write out
$\rho^{(N)}$ with respect to any orthonormal tensor-product basis
on ${\cal H}_d^{\otimes N}$, say
$|i_1\rangle|i_2\rangle\cdots|i_N\rangle$, where
$i_k\in\{1,2,\ldots,d\}$ for all $k\,$.  The joint state takes the
form
\begin{equation}
\rho^{(N)}=\sum_{i_1,\ldots,i_N;j_1,\ldots,j_N}
R^{(N)}_{i_1,\ldots,i_N;j_1,\ldots,j_N}\,
|i_1\rangle\cdots|i_N\rangle \langle j_1| \cdots\langle j_N|\;,
\end{equation}
where $R^{(N)}_{i_1,\ldots,i_N;j_1,\ldots,j_N}$ is the density
matrix in this representation.  What we demand is that for any
permutation $\pi$ of the set $\{1,\ldots,N\}$,
\begin{eqnarray}
\rho^{(N)}&=&\sum_{i_1,\ldots,i_N;j_1,\ldots,j_N}
R^{(N)}_{i_1,\ldots,i_N;j_1,\ldots,j_N}\,
|i_{\pi^{-1}(1)}\rangle\cdots|i_{\pi^{-1}(N)}\rangle
\langle j_{\pi^{-1}(1)}|\cdots\langle j_{\pi^{-1}(N)}|\nonumber\\
&=&\sum_{i_1,\ldots,i_N;j_1,\ldots,j_N}
R^{(N)}_{i_{\pi(1)},\ldots,i_{\pi(N)};j_{\pi(1)},\ldots,j_{\pi(N)}}\,
|i_1\rangle\cdots|i_N\rangle \langle j_1| \cdots\langle j_N|
\;,
\end{eqnarray}
which is equivalent to
\begin{equation}
R^{(N)}_{i_{\pi(1)},\ldots,i_{\pi(N)};j_{\pi(1)},\ldots,j_{\pi(N)}}
=R^{(N)}_{i_1,\ldots,i_N;j_1,\ldots,j_N}\;.
\end{equation}

The state $\rho^{(N)}$ is said to be {\it exchangeable\/} (or
infinitely exchangeable) if it is symmetric and if, for any $M>0$,
there is a symmetric state $\rho^{(N+M)}$ of $N+M$ systems such that
the marginal density operator for $N$ systems is $\rho^{(N)}$, i.e.,
\begin{equation}
\rho^{(N)} = \tr_M\,\rho^{(N+M)} \;,
\label{HoundDog}
\end{equation}
where the trace is taken over the additional $M$ systems.  In
explicit basis-dependent notation, this requirement is
\begin{equation}
\rho^{(N)}=
\!\!\sum_{i_1,\ldots,i_N;j_1,\ldots,j_N}
\!\!\left(\,\sum_{i_{N+1},\ldots,i_{N+M}}\!\!
R^{(N+M)}_{i_1,\ldots,i_N,i_{N+1},\ldots,i_{N+M};
j_1,\ldots,j_N,i_{N+1},\ldots,i_{N+M}}\right)\!
|i_1\rangle\cdots|i_N\rangle \langle j_1| \cdots\langle j_N|\;.
\end{equation}
In analogy to the classical case, an exchangeable density operator can
be thought of informally as the description of a subsystem of an
infinite sequence of systems whose order is irrelevant.

The precise statement of the quantum de Finetti representation
theorem~\cite{Hudson1976,Stormer1969} is that any exchangeable state
of $N$ systems can be written uniquely in the form
\begin{equation}
\rho^{(N)}=\int_{{\cal D}_d} P(\rho)\, \rho^{\otimes N}\, d\rho\;.
\label{eq-qdefinetti}
\end{equation}
Here $P(\rho)\ge0$ is normalized by
\begin{equation}
\int_{{\cal D}_d} P(\rho)\,d\rho=1\;,
\end{equation}
with $d\rho$ being a suitable measure on density operator space
${\cal D}_d$ [e.g., one could choose the standard flat measure
$d\rho=S^2dS\,d\Omega$ in the parametrization~(\ref{eq-rhoqubit})
for a spin-$1\over 2$ particle].  The upshot of the theorem, as
already advertised, is that it makes it possible to think of an
exchangeable quantum-state assignment {\it as if\/} it were a
probabilistic mixture characterized by a probability density
$P(\rho)$ for the product states $\rho^{\otimes N}$.

Just as in the classical case, both components of the definition
of exchangeability are crucial for arriving at the representation
theorem of Eq.~(\ref{eq-qdefinetti}).  The reason now, however, is
much more interesting than it was previously.  In the classical
case, extendibility was used solely to exclude anticorrelated
probability distributions. Here extendibility is necessary to
exclude the possibility of Bell inequality violations for
measurements on the separate systems. This is because the
assumption of symmetry alone for an $N$-party quantum system does
not exclude the possibility of quantum entanglement, and all
states that can be written as a mixture of product states---of
which Eq.~(\ref{eq-qdefinetti}) is an example---have no
entanglement~\cite{Bennett1996}. A very simple counterexample is
the Greenberger-Horne-Zeilinger state of three spin-$1\over2$
particles~\cite{Mermin1990},
\begin{equation}
|\mbox{GHZ}\rangle=\frac{1}{\sqrt{2}}\Big(|0\rangle|0\rangle|0\rangle+
|1\rangle|1\rangle|1\rangle\Big)\;,
\end{equation}
which is symmetric, but is not extendible to a symmetric state on
four systems.  This follows because the only states of four
particles that marginalize to a three-particle pure state, like
the GHZ state, are product states of the form
$|\mbox{GHZ}\rangle\langle\mbox{GHZ}|\otimes\rho$, where $\rho$ is
the state of the fourth particle; such states clearly cannot be
symmetric.  These considerations show that in order for the
proposed theorem to be valid, it must be the case that as $M$
increases in Eq.~(\ref{HoundDog}), the possibilities for
entanglement in the separate systems compensatingly
decrease~\cite{Koashi2000}.

\section{Proof of the quantum de Finetti theorem} \label{sec-proof}

To prove the quantum version of the de Finetti theorem, we rely on
the classical theorem as much as possible.  We start from an
exchangeable density operator $\rho^{(N)}$ defined on $N$ copies
of a system.  We bring the classical theorem to our aid by
imagining a sequence of identical quantum measurements on the
separate systems and considering the outcome probabilities they
would produce.  Because $\rho^{(N)}$ is assumed exchangeable, such
identical measurements give rise to an exchangeable probability
distribution for the outcomes.  The trick is to recover enough
information from the exchangeable statistics of these measurements
to characterize the exchangeable density operators.

With this in mind, the proof is expedited by making use of the
theory of generalized quantum measurements or positive
operator-valued measures (POVMs)~\cite{Peres1993a,Kraus1983}.  We
give a brief introduction to that theory. The common textbook
notion of a measurement---that is, a von Neumann measurement---is
that any laboratory procedure counting as an observation can be
identified with a Hermitian operator $O$ on the Hilbert space
${\cal H}_d$ of the system. Depending upon the presentation, the
measurement outcomes are identified either with the eigenvalues
$\mu_i$ or with a complete set of normalized eigenvectors
$|i\rangle$ for $O$. When the quantum state is $\rho$, the
probabilities for the various outcomes are computed from the
eigenprojectors $\Pi_i=|i\rangle\langle i|$ via the standard Born
rule,
\begin{equation}
p_i={\rm tr}\big(\rho\Pi_i\big) = \langle i|\rho|i\rangle\;.
\end{equation}
This rule gives a consistent probability assignment because the
eigenprojectors $\Pi_i$ are positive-semidefinite operators, which
makes the $p_i$ nonnegative, and because the projectors form a
resolution of the identity operator $I$,
\begin{equation}
\sum_{i=1}^d \Pi_i = I\;,
\end{equation}
which guarantees that $\sum_i p_i=1$.

POVMs generalize the textbook notion of measurement by distilling
the essential properties that make the Born rule work.  The
generalized notion of measurement is this:  {\it any\/} set ${\cal
E}=\{E_\alpha\}$ of positive-semidefinite operators  on ${\cal
H}_d$ that forms a resolution of the identity, i.e., that
satisfies
\begin{equation}
\langle\psi|E_\alpha|\psi\rangle\ge0\,,\quad\mbox{for all
$|\psi\rangle\in{\cal H}_d$}
\label{Hank}
\end{equation}
and
\begin{equation}
\sum_\alpha E_\alpha = I\;,
\label{Hannibal}
\end{equation}
corresponds to at least one laboratory procedure counting as a
measurement. The outcomes of the measurement are identified with the
indices $\alpha$, and the probabilities of those outcomes are
computed according to the generalized Born rule,
\begin{equation}
p_\alpha=\tr\big(\rho E_\alpha\big) \;.
\end{equation}
The set ${\cal E}$ is called a POVM, and the operators $E_\alpha$
are called POVM elements. Unlike von Neumann measurements, there
is no limitation on the number of values $\alpha$ can take, the
operators $E_\alpha$ need not be rank-1, and there is no
requirement that the $E_\alpha$ be mutually orthogonal.  This
definition has important content because the older notion of
measurement is simply too restrictive: there are laboratory
procedures that clearly should be called ``measurements,'' but
that cannot be expressed in terms of the von Neumann measurement
process alone.

One might wonder whether the existence of POVMs contradicts
everything taught about standard measurements in the traditional
graduate textbooks~\cite{QuantumClassics1} and the well-known
classics~\cite{QuantumClassics2}. Fortunately it does not. The
reason is that any POVM can be represented formally as a standard
measurement on an ancillary system that has interacted in the past
with the system of main interest.  Thus, in a certain sense, von
Neumann measurements capture everything that can be said about
quantum measurements \cite{Kraus1983}. A way to think about this
is that by learning something about the ancillary system through a
standard measurement, one in turn learns something about the
system of real interest. Indirect though this might seem, it can
be a very powerful technique, sometimes revealing information that
could not have been revealed otherwise~\cite{Holevo1973}.

For instance, by considering POVMs, one can consider measurements
with an outcome cardinality that exceeds the dimensionality of the
Hilbert space. What this means is that whereas the statistics of a
von Neumann measurement can only reveal information about the $d$
diagonal elements of a density operator $\rho$, through the
probabilities ${\rm tr}\big(\rho\Pi_i\big)$, the statistics of a
POVM generally can reveal things about the off-diagonal elements,
too. It is precisely this property that we take advantage of in our
proof of the quantum de Finetti theorem.

Our problem hinges on finding a special kind of POVM, one for
which any set of outcome probabilities specifies a unique
operator.  This boils down to a problem in pure linear algebra.
The space of operators on ${\cal H}_d$ is itself a linear vector
space of dimension $d^{\,2}$. The quantity ${\rm tr}(A^\dagger B)$
serves as an inner product on that space. If the POVM elements
$E_\alpha$ span the space of operators---there must be at least
$d^{\,2}$ POVM elements in the set---the measurement probabilities
$p_\alpha={\rm tr}\big(\rho E_\alpha\big)$---now thought of as
{\it projections\/} in the directions $E_\alpha$---are sufficient
to specify a unique operator $\rho$.  Two distinct density
operators $\rho$ and $\sigma$ must give rise to different
measurement statistics. Such measurements, which might be called
{\it informationally complete}, have been studied for some
time~\cite{Prugovecki1977}.

For our proof we need a slightly refined notion---that of a {\it
minimal\/} informationally complete measurement.  If an
informationally complete POVM has more than $d^{\,2}$ operators
$E_\alpha$, these operators form an overcomplete set.  This means
that given a set of outcome probabilities $p_\alpha$, there is
generally {\it no\/} operator $A$ that generates them according to
$p_\alpha={\rm tr}\big(AE_\alpha\bigr)$.  Our proof requires the
existence of such an operator, so we need a POVM that has
precisely $d^{\,2}$ linearly independent POVM elements $E_\alpha$.
Such a POVM has the minimal number of POVM elements to be
informationally complete.  Given a set of outcome probabilities
$p_\alpha$, there is a unique operator $A$ such that
$p_\alpha={\rm tr}\big(AE_\alpha\bigr)$, even though, as we
discuss below, $A$ is not guaranteed to be a density operator.

Do minimal informationally complete POVMs exist?  The answer is
yes. We give here a simple way to produce one, though there are
surely more elegant ways with greater symmetry. Start with a
complete orthonormal basis $|e_j\rangle$ on ${\cal H}_d$, and let
$\Gamma_{jk}=|e_j\rangle\langle e_k|$.  It is easy to check that
the following $d^{\,2}$ rank-1 projectors $\Pi_\alpha$ form a
linearly independent set.
\begin{enumerate}
\item For $\alpha=1,\ldots,d$, let
\begin{equation}
\Pi_\alpha \equiv \Gamma_{jj}\,,
\end{equation}
where $j$, too, runs over the values $1,\ldots,d$.

\item For $\alpha=d+1,\ldots,\frac{1}{2}d(d+1)$, let
\begin{equation}
\Pi_\alpha \equiv \Gamma^{(1)}_{jk} =
\frac{1}{2}\Big(|e_j\rangle+|e_k\rangle\Big)
\Big(\langle e_j|+\langle e_k|\Big)
=
\frac{1}{2}(\Gamma_{jj}+\Gamma_{kk}+\Gamma_{jk}+\Gamma_{kj})\;,
\end{equation}
where $j<k$.

\item Finally, for $\alpha= \frac{1}{2}d(d+1) + 1, \ldots,d^{\,2}$, let
\begin{equation}
\Pi_\alpha \equiv \Gamma^{(2)}_{jk}
= \frac{1}{2}\Big(|e_j\rangle+i|e_k\rangle\Big)
\Big(\langle e_j|-i\langle e_k |\Big)
=\frac{1}{2}(\Gamma_{jj}+\Gamma_{kk}-i\Gamma_{jk}+i\Gamma_{kj})\;,
\end{equation}
where again $j<k$.
\end{enumerate}
All that remains is to transform these (positive-semidefinite)
linearly independent operators $\Pi_\alpha$ into a proper POVM.
This can be done by considering the positive semidefinite operator
$G$ defined by
\begin{equation}
G=\sum_{\alpha=1}^{d^2}\Pi_\alpha\;.
\label{Herbert}
\end{equation}
It is straightforward to show that $\langle\psi|G|\psi\rangle>0$
for all $|\psi\rangle\ne0$, thus establishing that $G$ is positive
definite (i.e., Hermitian with positive eigenvalues) and hence
invertible.  Applying the (invertible) linear transformation
$X\rightarrow\, G^{-1/2}XG^{-1/2}$ to Eq.~(\ref{Herbert}), we find
a valid decomposition of the identity,
\begin{equation}
I=\sum_{\alpha=1}^{d^2}G^{-1/2}\Pi_\alpha G^{-1/2}\;.
\end{equation}
The operators
\begin{equation}
E_\alpha=G^{-1/2}\Pi_\alpha G^{-1/2}
\end{equation}
satisfy the conditions of a POVM, Eqs.~(\ref{Hank}) and
(\ref{Hannibal}), and moreover, they retain the rank and
linear independence of the original $\Pi_\alpha$.

With this generalized measurement (or any other one like it), we
can return to the main line of proof.  Recall we assumed that we
captured our state of knowledge by an exchangeable density
operator $\rho^{(N)}$. Consequently, repeated application of the
(imagined) measurement $\cal E$ must give rise to an exchangeable
probability distribution over the $N$ random variables
$\alpha_n\in\{1,2,\ldots,d^{\,2}\}$, $n=1,\ldots,N$.   We now
analyze these probabilities.

Quantum mechanically, it is valid to think of the $N$ repeated
measurements of $\cal E$ as a single measurement on the Hilbert
space ${\cal H}_d^{\otimes N}\equiv {\cal
H}_d\otimes\cdots\otimes{\cal H}_d$. This measurement, which we
denote ${\cal E}^{\otimes N}$, consists of $d^{\,2N}$ POVM
elements of the form $E_{\alpha_1}\otimes\cdots\otimes
E_{\alpha_N}$. The probability of any particular outcome sequence
of length $N$, namely
$\bbox{\alpha}\equiv(\alpha_1,\ldots,\alpha_N)$, is given by the
standard quantum rule,
\begin{equation}
p^{(N)}(\bbox{\alpha})={\rm
tr}\big(\,\rho^{(N)}\,E_{\alpha_1}\otimes\cdots\otimes
E_{\alpha_N}\big)\;.
\label{Humphrey}
\end{equation}
Because the distribution $p^{(N)}(\bbox{\alpha})$ is exchangeable,
we have by the classical de Finetti theorem [see
Eq.~(\ref{eq-repr})] that there exists a unique probability
density $P(\vec{p})$ on ${\cal S}_{d^2}$ such that
\begin{equation}
p^{(N)}(\bbox{\alpha})= \int_{{\cal S}_{d^2}} P(\vec{p})\,
p_{\alpha_1} p_{\alpha_2}\cdots p_{\alpha_N}\,d\vec{p}\;.
\label{Helmut}
\end{equation}

It should now begin to be apparent why we chose to imagine a
measurement $\cal E$ consisting of precisely $d^{\,2}$ linearly
independent elements. This allows us to assert the existence of a
{\it unique\/} operator $A_{\vec{p}}$ on ${\cal H}_d$
corresponding to each point $\vec{p}$ in the domain of the
integral.  The ultimate goal here is to turn Eqs.~(\ref{Humphrey})
and (\ref{Helmut}) into a single operator equation.

With that in mind, let us define $A_{\vec{p}}$ as the unique
operator satisfying the following $d^{\,2}$ linear equations:
\begin{equation}
{\rm tr}\big(A_{\vec{p}}E_\alpha\big)=
p_\alpha\;,\quad\quad\alpha=1,\ldots,d^{\,2}\;.
\label{Hamish}
\end{equation}
Inserting this definition into Eq.~(\ref{Helmut}) and manipulating it
according to the algebraic rules of tensor products---namely
$(A\otimes B)(C\otimes D)=AC\otimes BD$ and ${\rm tr}(A\otimes
B)=({\rm tr}A)({\rm tr}B)$---we see that
\begin{eqnarray}
p^{(N)}(\bbox{\alpha})
&=&
\int_{{\cal S}_{d^2}} P(\vec{p})\,{\rm
tr}\big(A_{\vec{p}}E_{\alpha_1}\big) \cdots {\rm
tr}\big(A_{\vec{p}}E_{\alpha_N}\big)\,d\vec{p}
\nonumber\\
&=&
\int_{{\cal S}_{d^2}} P(\vec{p})\,{\rm
tr}\big(A_{\vec{p}}E_{\alpha_1}\otimes \cdots\otimes
A_{\vec{p}}E_{\alpha_N}\big)\,d\vec{p}
\nonumber\\
&=&
\int_{{\cal S}_{d^2}} P(\vec{p})\,{\rm tr}\big[A_{\vec{p}}^{\otimes N}
\,
(E_{\alpha_1}\otimes\cdots\otimes E_{\alpha_N})\big]\,d\vec{p}\;.
\end{eqnarray}
If we further use the linearity of the trace, we can write the
same expression as
\begin{equation}
p^{(N)}(\bbox{\alpha})={\rm tr}\!\left[\left(\int_{{\cal S}_{d^2}}
P(\vec{p})\,A_{\vec{p}}^{\otimes n}
\,\,d\vec{p}\right)E_{\alpha_1}\otimes\cdots\otimes
E_{\alpha_N}\right].
\label{Hugo}
\end{equation}

The identity between Eqs.~(\ref{Humphrey}) and (\ref{Hugo}) must hold
for all sequences $\bbox{\alpha}$.  It follows that
\begin{equation}
\rho^{(N)}=\int_{{\cal S}_{d^2}} P(\vec{p})\,A_{\vec{p}}^{\otimes N}
\,\,d\vec{p}\;.
\label{Howard}
\end{equation}
This is because the operators $E_{\alpha_1}\otimes\cdots\otimes
E_{\alpha_N}$ form a complete basis for the vector space of
operators on ${\cal H}_d^{\otimes N}$.

Equation~(\ref{Howard}) already looks very much like our sought
after goal, but we are not there quite yet.  At this stage one has no
right to assert that the $A_{\vec{p}}$ are density operators. Indeed
they generally are not:  the integral~(\ref{Helmut}) ranges over
some points $\vec{p}$ in ${\cal S}_{d^{\,2}}$ that cannot be
generated by applying the measurement $\cal E$ to {\it any\/}
quantum state.  Hence some of the $A_{\vec{p}}$ in the integral
representation are ostensibly nonphysical.  An example might be
helpful. Consider any four spin-$1\over2$ pure states $|\vec
n_\alpha\rangle$ on ${\cal H}_2$ for which the vectors $\vec
n_\alpha$ in the Bloch-sphere representation~(\ref{eq-poincare}) are
the vertices of a regular tetrahedron.  One can check that the
elements $E_\alpha=\frac{1}{2}|\vec n_\alpha\rangle\langle\vec
n_\alpha|$ comprise a minimal informationally complete POVM. For
this POVM, because of the factor $\frac{1}{2}$ in front of each
projector, it is always the case that $p_\alpha={\rm tr}(\rho
E_\alpha)\le\frac{1}{2}$. Therefore, this measurement simply cannot
generate a probability distribution like
$\vec{p}=\big(\frac{3}{4},\frac{1}{8},\frac{1}{16},\frac{1}{16}\big)$,
which is nevertheless in the domain of the integral in
Eq.~(\ref{Helmut}).

The solution to this conundrum is provided by the overall requirement
that $\rho^{(N)}$ be a valid density operator.  This requirement
places a significantly more stringent constraint on the distribution
$P(\vec{p})$ than was the case in the classical representation
theorem.  In particular, it must be the case that $P(\vec{p})$
vanishes whenever the corresponding $A_{\vec{p}}$ is not a proper
density operator.  Let us move toward showing that.

We first need to delineate two properties of the operators
$A_{\vec{p}}$.  One is that they are Hermitian. The argument is
simply
\begin{equation}
{\rm tr}\big(E_\alpha A_{\vec{p}}^\dagger\big) = {\rm
tr}\!\left[\big(A_{\vec{p}}E_\alpha\big)^\dagger\right] = \big[{\rm
tr}\big(A_{\vec{p}}E_\alpha\big)\big]^* = {\rm
tr}\big(A_{\vec{p}}E_\alpha\big)\;,
\end{equation}
where the last step follows from Eq.~(\ref{Hamish}).  Because the
$E_\alpha$ are a complete set of linearly independent operators,
it follows that $A_{\vec{p}}^\dagger=A_{\vec{p}}$.  The second
property tells us something about the eigenvalues of
$A_{\vec{p}}$:
\begin{equation}
1=\sum_\alpha p_\alpha={\rm tr}\!\left(A_{\vec{p}}\sum_\alpha
E_\alpha\right)={\rm tr}A_{\vec{p}}\;.
\label{HepPlease}
\end{equation}
In other words the (real) eigenvalues of $A_{\vec{p}}$ must sum to
unity.

We now show that these two facts go together to imply that if
there are any nonphysical $A_{\vec{p}}$ with positive weight
$P(\vec{p})$ in Eq.~(\ref{Howard}), then one can find a
measurement for which $\rho^{(N)}$ produces illegal
``probabilities'' for sufficiently large $N$.  For instance, take
a particular $A_{\vec{q}}$ in Eq.~(\ref{Howard}) that has at least
one negative eigenvalue $-\lambda<0$.  Let $|\psi\rangle$ be a
normalized eigenvector corresponding to that eigenvalue and
consider the binary-valued POVM consisting of the elements
$\widetilde{\Pi}=|\psi\rangle\langle\psi|$ and
$\Pi=I-\widetilde{\Pi}$.  Since ${\rm
tr}\big(A_{\vec{q}}\widetilde{\Pi}\big)=-\lambda<0$, it is true by
Eq.~(\ref{HepPlease}) that ${\rm
tr}\big(A_{\vec{q}}\Pi\big)=1+\lambda >1$. Consider repeating this
measurement over and over. In particular, let us tabulate the
probability of getting outcome $\Pi$ for every single trial to the
exclusion of all other outcomes.

The gist of the contradiction is most easily seen by {\it imagining\/}
that Eq.~(\ref{Howard}) is really a discrete sum:
\begin{equation}
\rho^{(N)}= P(\vec{q})\,A_{\vec{q}}^{\otimes
N}+\sum_{\vec{p}\ne\vec{q}}P(\vec{p})\,A_{\vec{p}}^{\otimes N}\;.
\end{equation}
The probability of $N$ occurrences of the outcome $\Pi$ is thus
\begin{eqnarray}
{\rm tr}\big(\rho^{(N)}\Pi^{\otimes N}\big)
&=&
P(\vec{q})\,{\rm tr}(A_{\vec{q}}^{\otimes N}\Pi^{\otimes N})
+\sum_{\vec{p}\ne\vec{q}}P(\vec{p})\,{\rm tr}(A_{\vec{p}}^{\otimes
N}\Pi^{\otimes N})
\nonumber\\
&=&
P(\vec{q})\,[{\rm tr}(A_{\vec{q}}\Pi)]^N
+\sum_{\vec{p}\ne\vec{q}}P(\vec{p})\,[{\rm tr}(A_{\vec{p}}\Pi)]^N
\nonumber\\
&=&
P(\vec{q})(1+\lambda)^N +\sum_{\vec{p}\ne\vec{q}}P(\vec{p})\,[{\rm
tr}(A_{\vec{p}}\Pi)]^N\;.
\label{Hanna}
\end{eqnarray}
There are no assurances in general that the rightmost term in
Eq.~(\ref{Hanna}) is positive, but if $N$ is an even number it
must be.  It follows that if $P(\vec{q})\ge0$, for sufficiently
large {\it even\/} $N$,
\begin{equation}
{\rm tr}\big(\rho^{(N)}\Pi^{\otimes N}\big)>1\;,
\label{BigBoy}
\end{equation}
contradicting the assumption that it should always be a probability.

All we need to do now is transcribe the argument leading to
Eq.~(\ref{BigBoy}) to the general integral case of
Eq.~(\ref{Howard}). Note that by Eq.~(\ref{Hamish}), the quantity
${\rm tr}\big(A_{\vec{p}}\Pi\big)$ is a (linear) continuous
function of the parameter $\vec{p}$.  Therefore, for any
$\epsilon>0$, there exists a $\delta>0$ such that $\big|{\rm
tr}\big(A_{\vec{p}}\Pi\big)- {\rm
tr}\big(A_{\vec{q}}\Pi\big)\big|\le\epsilon$ whenever
$|\vec{p}-\vec{q}|\le\delta$, i.e., whenever $\vec{p}$ is
contained within an open ball $B_\delta(\vec{q})$ centered at
$\vec{q}$.  Choose $\epsilon<\lambda$, and define
$\overline{B}_\delta$ to be the intersection of
$B_\delta(\vec{q})$ with the probability simplex. For $\vec{p}\in
\overline{B}_\delta$, it follows that
\begin{equation}
{\rm tr}\big(A_{\vec{p}}\Pi\big)\ge 1+\lambda-\epsilon>1\;.
\end{equation}
If we consider an $N$ that is even, $\big[{\rm
tr}\big(A_{\vec{p}}\Pi\big)\big]^N$ is nonnegative in all of
${\cal S}_{d^2}$, and we have that the probability of the outcome
$\Pi^{\otimes N}$ satisfies
\begin{eqnarray}
{\rm tr}\big(\rho^{(N)}\Pi^{\otimes N}\big)
&=&
\int_{{\cal S}_{d^2}} P(\vec{p})\,\big[{\rm
tr}\big(A_{\vec{p}}\Pi\big)\big]^N \,d\vec{p}\;
\nonumber\\
&=&
\int_{{\cal S}_{d^2}-\overline{B}_\delta} P(\vec{p})\,\big[{\rm
tr}\big(A_{\vec{p}}\Pi\big)\big]^N \,d\vec{p}
\; +\, \int_{\overline{B}_\delta} P(\vec{p})\,\big[{\rm
tr}\big(A_{\vec{p}}\Pi\big)\big]^N \,d\vec{p}
\nonumber\\
&\ge&
\int_{\overline{B}_\delta} P(\vec{p})\,\big[{\rm
tr}\big(A_{\vec{p}}\Pi\big)\big]^N \,d\vec{p}
\nonumber\\
&\ge&
(1+\lambda-\epsilon)^N \int_{\overline{B}_\delta}
P(\vec{p})\,d\vec{p}\;.
\label{Homer}
\end{eqnarray}
Unless
\begin{equation}
\int_{\overline{B}_\delta} P(\vec{p})\,d\vec{p}=0\;,
\end{equation}
the lower bound (\ref{Homer}) for the probability of the outcome
$\Pi^{\otimes N}$ becomes arbitrarily large as
$N\rightarrow\infty$.  Thus we conclude that the requirement that
$\rho^{(N)}$ be a proper density operator constrains $P(\vec{p})$
to vanish almost everywhere in $\overline{B}_\delta$ and,
consequently, to vanish almost everywhere that $A_{\vec{p}}$ is
not a physical state.

Using Eq.~(\ref{Hamish}), we can trivially transform the integral
representation (\ref{Howard}) to one directly over the convex set
of density operators ${\cal D}_d$ and be left with the following
statement. Under the sole assumption that the density operator
$\rho^{(N)}$ is exchangeable, there exists a unique probability
density $P(\rho)$ such that
\begin{equation}
\rho^{(N)}=\int_{{\cal D}_d} P(\rho)\, \rho^{\otimes N}\, d\rho\;.
\end{equation}
This concludes the proof of the quantum de Finetti representation
theorem.

\section{Outlook} \label{sec-outlook}

Since the analysis in the previous sections concerned {\it only\/}
the case of quantum-state tomography, we certainly have not
written the last word on unknown quantum states in the sense
advocated in Sec.~\ref{sec-intro}. There are clearly other
examples that need separate analyses. For instance, the use of
unknown states in quantum teleportation~\cite{Bennett1993}---where
a {\it single\/} realization of an unknown state is ``teleported''
with the aid of previously distributed quantum entanglement and a
classical side channel---has not been touched upon. The quantum de
Finetti theorem, therefore, is not the end of the road for
detailing implications of an information-based interpretation of
quantum mechanics. What is important, we believe, is that taking
the time to think carefully about the referents of various states
in a problem can lead to insights into the structure of quantum
mechanics that cannot be found by other means.

For instance, one might ask, ``Was this theorem not inevitable?''
After all, is it not already well established that quantum theory
is, in some sense, just a noncommutative generalization of
probability theory?  Should not all the main theorems in classical
probability theory carry over to the quantum
case~\cite{CommentAccardi}?  One can be skeptical in this way, of
course, but then one will miss a large part of the point.  There
are any number of noncommutative generalizations to probability
theory that one can concoct~\cite{Hiai2000}.  The deeper issue is,
what is it in the natural world that forces quantum theory to the
particular noncommutative structure it actually
has~\cite{Wheeler2000}? It is not a foregone conclusion, for
instance, that every theory has a de Finetti representation
theorem within it.

Some insight in this regard can be gained by considering very
simple modifications of quantum theory.  To give a concrete
example, let us take the case of real-Hilbert-space quantum
mechanics.  This theory is the same as ordinary quantum mechanics
in all aspects {\it except\/} that the Hilbert spaces are defined
over the field of real numbers rather than the complex numbers.
It turns out that this is a case where the quantum de Finetti
theorem fails.  Let us start to explain why by first describing
how the particular proof technique used above loses validity in
the new context.

In order to specify uniquely a Hermitian operator $\rho^{(N)}$ in
going from Eq.~(\ref{Hugo}) to (\ref{Howard}), the proof made
central use of the fact that a complete basis
$\{E_{1},\ldots,E_{d^{\,2}}\}$ for the vector space of operators
on ${\cal H}_d$ can be used to generate a complete basis for the
operators on ${\cal H}_d^{\otimes N}$---one just need take the
$d^{\,2N}$ operators of the form
$E_{\alpha_1}\!\otimes\cdots\otimes E_{\alpha_N}$,
$1\le\alpha_j\le d^{\,2}$.  (All we actually needed was that a
basis for the real vector space of Hermitian operators on ${\cal
H}_d$ can be used to generate a basis for the real vector space of
Hermitian operators on ${\cal H}_d^{\otimes N}$, but since the
vector space of all operators is the complexification of the real
vector space of Hermitian operators, this seemingly weaker
requirement is, in fact, no different.)  This technique works
because the dimension of the space of $d^N\!\times d^N$ matrices
is $(d^{\,2})^N$, the $N$th power of the dimension of the space
of $d\times d$ matrices.

This technique does not carry over to real Hilbert spaces. In a
real Hilbert space, states and POVM elements are represented by
real symmetric matrices.  The dimension of the vector space of
real symmetric matrices acting on a $d$-dimensional real Hilbert
space is ${1\over2}d(d+1)$, this then being the number of elements
in an minimal informationally complete POVM.  The task in going
from Eq.~(\ref{Hugo}) to (\ref{Howard}) would be to specify the
real matrix $\rho^{(N)}$.  When $N\ge2$, however, the dimension of
the space of $d^N\!\times d^N$ real symmetric matrices is strictly
greater than the $N$th power of the dimension of the space of
$d\times d$ real symmetric matrices, i.e.,
\begin{equation}
{1\over2}d^{N}(d^N+1)>\left({1\over2}d(d+1)\right)^{\! N}\;.
\end{equation}
Hence, specifying Eq.~(\ref{Hugo}) for all outcome sequences
$\bbox{\alpha}=(\alpha_1,\ldots,\alpha_N)$ is not sufficient to
specify a single operator $\rho^{(N)}$.  This line of reasoning
indicates that the particular {\it proof\/} of the quantum de
Finetti theorem presented in Sec.~\ref{sec-proof} fails for real
Hilbert spaces, but it does not establish that the theorem itself
fails.  The main point of this discussion is that it draws
attention to the crucial difference between real-Hilbert-space and
complex-Hilbert-space quantum mechanics---a fact emphasized
previously by Araki~\cite{Araki1980} and
Wootters~\cite{Wootters1990}.

To show that the theorem fails, we need a counterexample. One such
example is provided by the $N$-system state
\begin{equation}
\rho^{(N)}={1\over2}\,\rho_+^{\otimes N} + {1\over2}\,\rho_-^{\otimes N} \;,
\label{eq-real}
\end{equation}
where
\begin{equation}
\rho_+={1\over2}(I+\sigma_2)\qquad\mbox{and}
\qquad\rho_-={1\over2}(I-\sigma_2)\;,
\end{equation}
and where $\sigma_2$ was defined in Eq.~(\ref{eq-pauli}).
In complex-Hilbert-space quantum mechanics, this is clearly a
valid density operator: It corresponds to an equally weighted
mixture of $N$ spin-up particles and $N$ spin-down particles in
the $y$ direction.  The state $\rho^{(N)}$ is clearly
exchangeable, and the decomposition in Eq.~(\ref{eq-real}) is
unique according to the quantum de Finetti theorem.

Now consider $\rho^{(N)}$ as an operator in real-Hilbert-space
quantum mechanics.  Despite the apparent use of the imaginary number
$i$ in the $\sigma_2$ operator, $\rho^{(N)}$ remains a valid quantum
state.  This is because, upon expanding the right-hand side of
Eq.~(\ref{eq-real}), all the terms with an odd number of $\sigma_2$
operators cancel away.  Yet, even though it is an exchangeable
density operator, it cannot be written in de Finetti form of
Eq.~(\ref{eq-qdefinetti}) using only real symmetric operators. This
follows because Eq.~(\ref{eq-real}), the unique de Finetti form,
contains $\sigma_2$, which is an antisymmetric operator and cannot
be written in terms of symmetric operators.  Hence the de Finetti
representation theorem does not hold in real-Hilbert-space quantum
mechanics.

Similar considerations show that in quaternionic quantum mechanics
(a theory again precisely the same as ordinary quantum mechanics
except that it uses Hilbert spaces over the quaternionic
field~\cite{Adler1995}), the connection between exchangeable density
operators and decompositions of the de Finetti
form~(\ref{eq-qdefinetti}) breaks down.  The failure mode is,
however, even more disturbing than for real Hilbert spaces. In
quaternionic quantum mechanics, most operators of the de Finetti
form~(\ref{eq-qdefinetti}) do not correspond to valid quaternionic
quantum states, even though the states $\rho$ in the integral are
valid quaternionic states.  The reason is that tensor products of
quaternionic Hermitian operators are not necessarily Hermitian.

In classical probability theory, exchangeability characterizes
those situations where the only data relevant for updating a
probability distribution are frequency data, i.e., the numbers
$n_j$ in Eq.~(\ref{eq-repr}) which tell how often the result $j$
occurred. The quantum de Finetti representation shows that the
same is true in quantum mechanics:  Frequency data (with respect
to a sufficiently robust measurement) are sufficient for updating
an exchangeable state to the point where nothing more can be
learned from sequential measurements; that is, one obtains a
convergence of the form~(\ref{HannibalLecter}), so that ultimately
any further measurements on the individual systems are
statistically independent. That there is no quantum de Finetti
theorem in real Hilbert space means that there are fundamental
differences between real and complex Hilbert spaces with respect
to learning from measurement results.  The ultimate reason for
this is that in ordinary, complex-Hilbert-space quantum mechanics,
exchangeability implies separability, i.e., the absence of
entanglement.  This follows directly from the quantum de Finetti
theorem, because states of the form Eq.~(\ref{eq-qdefinetti}) are
not entangled.  This implication does not carry over to real
Hilbert spaces.  By the same reasoning used to show that the de
Finetti theorem itself fails, the state in Eq.~(\ref{eq-real})
cannot be written as {\it any\/} mixture of real product states.
Interpreted as a state in real Hilbert space, the state in
Eq.~(\ref{eq-real}) is thus not separable, but
entangled~\cite{Caves2000}. In a real Hilbert space, exchangeable
states can be entangled and local measurements cannot reveal that.

Beyond these conceptual points, we also believe that the technical
methods exhibited here might be of interest in the practical arena.
Recently there has been a large literature on which classes of
measurements have various advantages for tomographic
purposes~\cite{QuorumLump,QuorumOld}.  To our knowledge, the present
work is the only one to consider tomographic reconstruction based
upon minimal informationally complete POVMs. One can imagine several
advantages to this approach via the fact that such POVMs with
rank-one elements are automatically extreme points in the convex set
of all measurements~\cite{Fujiwara1998}.

Furthermore, the classical de Finetti theorem is only the tip of an
iceberg with respect to general questions in statistics to do with
exchangeability and various generalizations of the
concept~\cite{Aldous1985}.  One should expect no less of quantum
exchangeability studies.  In particular here, we are thinking of
things like the question of representation theorems for finitely
exchangeable distributions~\cite{Jaynes1986,Diaconis1977}.  Just as
our method for proving the quantum de Finetti theorem was able to
rely heavily on the the classical theorem, so one might expect
similar benefits from the classical results in the case of quantum
finite exchangeability---although there will certainly be new
aspects to the quantum case due to the possibility of entanglement
in finite exchangeable states.  A practical application of such
representation theorems could be their potential to contribute to the
solution of some outstanding problems in constructing security
proofs for various quantum key distribution
schemes~\cite{Gottesman2000}.

In general, our effort in the present paper forms part of a larger
program to promote a consistent information-based interpretation of
quantum mechanics and to delineate its consequences.  We find it
encouraging that the fruits of this effort may not be restricted
solely to an improved understanding of quantum mechanics, but also
possess the potential to contribute to practical applications.

\acknowledgments

We thank Ben Schumacher for discussions.  Most of this work was
carried out through the hospitality of the Benasque Center for
Physics, Benasque, Spain during their program on Progress in Quantum
Computing, Cryptography and Communication, 5--25 July 1998, and the
hospitality of the Isaac Newton Institute for Mathematical Sciences,
Cambridge, England during their Workshop on Complexity, Computation
and the Physics of Information, June--July 1999.  CMC was supported
in part by Office of Naval Research Grant No.~N00014-93-1-0116.

\appendix

\section{Proof of the Classical de Finetti Theorem}

In this Appendix we reprise the admirably simple proof of the
classical de Finetti representation theorem given by Heath and
Sudderth \cite{Heath1976} for the case of binary variables.

Suppose we have an exchangeable probability assignment for $M$ binary
random variables, $x_1,x_2,\ldots,x_M$, taking on the values 0 and
1.  Let $p(n,N)$, $N\le M$, be the probability for $n$ 1s in $N$
trials.  Exchangeability guarantees that
\begin{equation}
p(n,N)= {N\choose n} p(x_1=1,\ldots,x_n=1,x_{n+1}=0,\ldots,x_N=0)\;.
\end{equation}
We can condition the probability on the right on the occurrence of $m$
1s in all $M$ trials:
\begin{equation}
p(n,N)={N\choose n} \sum_{m=0}^M
p(x_1=1,\ldots,x_n=1,x_{n+1}=0,\ldots,x_N=0\mid m,M) p(m,M)\;.
\end{equation}
Given $m$ 1s in $M$ trials, exchangeability guarantees that the
$\displaystyle{{M\choose m}}$ sequences are equally likely.  Thus
the situation is identical to drawing from an urn with $m$ 1s on
$M$ balls, and we have that
\begin{eqnarray}
&& p(x_1=1,\ldots,x_n=1,x_{n+1}=0,\ldots,x_N=0\mid m,M) \nonumber
\\
&&\hphantom{p(x_1=1,}{}= {m\over M} {m-1\over M-1} \cdots {m-(n-1)\over
M-(n-1)} {M-m\over M-n} {M-m-1\over M-n-1} \cdots {M-m-(N-n-1)\over
M-(N-1)} \nonumber \\
&&\hphantom{p(x_1=1,}{}= {(m)_n(M-m)_{N-n}\over(M)_N} \;,
\end{eqnarray}
where
\begin{equation}
(r)_q\equiv\prod_{j=0}^{q-1}(r-j)=r(r-1)\cdots(r-q+1)={r!\over(r-q)!}\;.
\end{equation}
The result is that
\begin{equation}
p(n,N)={N\choose n} \sum_{m=0}^M {(m)_n(M-m)_{N-n}\over(M)_N} p(m,M)\;.
\end{equation}

What remains is to take the limit $M\rightarrow\infty$, which we
can do because of the extendibility property of exchangeable
probabilities.  We can write $p(n,N)$ as an integral
\begin{equation}
p(n,N)={N\choose n} \int_0^1
{(zM)_n\bigl((1-z)M\bigr)_{N-n}\over(M)_N} P_M(z)\,dz\;,
\end{equation}
where
\begin{equation}
P_M(z)=
\sum_{m=0}^M p(zM,M)\delta (z-m/M)
\end{equation}
is a distribution concentrated at the $M$-trial frequencies $m/M$.
In the limit $M\rightarrow\infty$, $P_M(z)$ converges to a
continuous distribution $P_\infty(z)$, and the other terms in the
integrand go to $z^n(1-z)^{N-n}$, giving
\begin{equation}
p(n,N)={N\choose n} \int_0^1 z^n(1-z)^{N-n} P_\infty(z)\,dz \;.
\end{equation}
We have demonstrated the classical de Finetti representation
theorem for binary variables: If $p(n,N)$ is part of an infinite
exchangeable sequence, then it has a de Finetti representation in
terms of a ``probability on probabilities'' $P_\infty(z)$.  The
proof can readily be extended to nonbinary variables.


\begin{thebibliography}{10}
\bibitem{BerkeleyRhyme}
Placing the player here would be about as respectable/responsible as
George Berkeley's famous patch to his philosophical system of
idealism.  The difficulty is captured engagingly by a limerick of
Ronald Knox and its anonymous reply:
\begin{quotation}
\footnotesize
\noindent There was a young man who said, ``God \\
Must think it exceedingly odd \\
If he finds that this tree \\
Continues to be \\
\smallskip When there's no one about in the Quad.''
\end{quotation}
\medskip\noindent{\footnotesize REPLY}
\begin{quotation}
\footnotesize
\noindent Dear Sir: \\
Your astonishment's odd: \\
I am always about in the Quad. \\
And that's why the tree \\
Will continue to be, \\
Since observed by \\
\indent \indent \indent Yours faithfully, \\
\indent \indent \indent God.
\end{quotation}
\end{thebibliography}

\end{document}